\documentclass[pra,twocolumn,showpacs,amsmath,amstex,amssymb,citeautoscript]{revtex4-2}

\usepackage[T1]{fontenc}
\usepackage[utf8]{inputenc}
\usepackage{lipsum}
\usepackage{amsmath}
\usepackage{amssymb}
\usepackage{bbm}
\usepackage{braket}
\usepackage{xcolor}
\usepackage{pifont}
\usepackage[mathscr]{euscript}
\usepackage[shortlabels]{enumitem}
\usepackage[justification=justified]{subcaption}
\usepackage{graphicx}
\usepackage{lipsum}
\allowdisplaybreaks
\usepackage{float}
\usepackage{graphicx}
\usepackage{dsfont}
\usepackage{comment}
\usepackage[colorlinks=true]{hyperref}  
\hypersetup{
    bookmarks=true,         
    unicode=false,          
    pdftoolbar=true,        
    pdfmenubar=true,        
    pdffitwindow=false,     
    pdfstartview={FitH},    
    pdftitle={Flstar},    
    pdfauthor={Christos, Sachdev},     
    pdfsubject={},   
    pdfcreator={},   
    pdfproducer={}, 
    pdfkeywords={} {} {}, 
    pdfnewwindow=true,      
    colorlinks=true,       
    linkcolor=blue, 
    citecolor=blue,        
    filecolor=magenta,      
    urlcolor=blue           
} 

\usepackage{pdfpages} 
\usepackage{pgffor} 

\makeatletter
\AtBeginDocument{\let\LS@rot\@undefined}
\makeatother

\def\supplementfilename{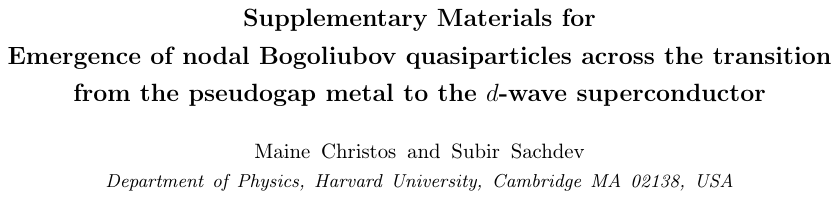}

\pdfximage{\supplementfilename}
\def\numbersupplementpages{\the\pdflastximagepages}

\newif\ifarXiv
\arXivtrue

\renewcommand{\vec}[1]{\boldsymbol{#1}}

\definecolor{wrongultramarine}{rgb}{1,0.5,0}

\begin{document}

\title{Emergence of nodal Bogoliubov quasiparticles across the transition\\ from the pseudogap metal to the $d$-wave superconductor}

\begin{abstract} 
We model the pseudogap state of the hole- and electron-doped cuprates as a metal with hole and/or electron pocket Fermi surfaces. In the absence of long-range antiferromagnetism, such Fermi surfaces violate the Luttinger requirement of enclosing the same area as free electrons at the same density. Using the Ancilla theory of such a pseudogap state, we describe the onset of conventional $d$-wave superconductivity by the condensation of a charge $e$ Higgs boson transforming as a fundamental under the emergent SU(2) gauge symmetry of a background $\pi$-flux spin liquid. In all cases, we find that the $d$-wave superconductor has gapless Bogoliubov quasiparticles at 4 nodal points on the Brillouin zone diagonals with significant velocity anisotropy, just as in the BCS state. This includes the case of the electron-doped pseudogap metal with only electron pockets centered at wavevectors $(\pi, 0)$, $(0,\pi)$, and an electronic gap along the zone diagonals. Remarkably, in this case too, gapless nodal Bogoliubov quasiparticles emerge within the gap at 4 points along the zone diagonals upon the onset of superconductivity. 
\end{abstract}

\author{Maine Christos}
\email{mainechristos@g.harvard.edu}
\thanks{Corresponding author.}
\affiliation{Department of Physics, Harvard University, Cambridge MA 02138, USA}

\author{Subir Sachdev}
\email{sachdev@g.harvard.edu}
\thanks{Corresponding author.}
\affiliation{Department of Physics, Harvard University, Cambridge MA 02138, USA}

\maketitle
\section{Introduction}
The remarkable phase diagram of the cuprates \cite{Keimer,Proust_2019,Damascelli_2003,Vishik_2010} has inspired an outpouring of theoretical and experimental work to explain their highly exotic phenomenology. Among the most extensively studied phases are the pseudogap metal, a phase characterized by a carrier density which deviates from the expectations required by Luttinger's theorem for a conventional Fermi liquid \cite{ShenRMP06,PhysRevLett.107.047003,doi:10.1126/science.1103627} and $d$-wave superconductivity which sets in at lower temperatures as an instability of the pseudogap phase \cite{RevModPhys.72.969,Ding_Yokoya}.

However, despite years of theoretical and experimental progress, a clear understanding of how superconductivity emerges from the experimentally observed pseudogap parent state and its associated small Fermi surface or Fermi arcs remains lacking. This work seeks to provide some basic answers as to what the experimental signatures of the transition from the pseudogap to superconductivity are. Although the pseudogap phase and its associated violation of Luttinger's theorem has been studied most extensively in the hole-doped cuprates, photo-emission experiments in the electron-doped cuprates have provided evidence for a reconstructed Fermi surface at dopings where long range antiferromagnetic order is believed to be absent \cite{Helm10,Shen19,Shen_2023}. The pairing in the electron-doped case is also believed to be $d$-wave \cite{Horio19}. We will therefore separately consider both the electron-doped and hole-doped cases in this work.

A number of works \cite{ID03,Stanescu_2006,Berthod_2006,Yang_2006,Sakai_2009,Robinson_2019,Skolimowski_2022,Fabrizio23,Giorgio23,Phillips23,Lucila23} have developed a model of the pseudogap metal in which the violation of the Luttinger theorem is associated with zeros of the electron Green's functions. Here, we view these zeros as a signal of the existence of an additional sector of neutral spinon excitations, which are required by non-perturbative extensions of the Luttinger theorem \cite{MO00,Senthil_2004}. As we will see below (and has also been argued earlier \cite{Chatterjee_2016}), a full treatment of the spinon sector is essential in understanding how the nodal Bogoliubov quasiparticles in the $d$-wave superconductor emerge from the pseudogap metal.

We will employ a theory \cite{Christos:2023oru} of the pseudogap metal with fermionic spinons coupled to to an SU(2) gauge field moving in a background of $\pi$-flux \cite{PhysRevB.37.3774,Wen_1996,Wen_2002,lee2004doping}. The fermionic spinons are coupled to physical electrons which carry the doping via a charge $e$ boson $B$ \cite{Wen_1996,Lee_1998,lee2004doping} which transforms under the same gauge SU(2) symmetry as the spinons. In the hole-doped case, due to the presence of the spin liquid, the normal state electron Fermi surface will have pockets associated with hole density $p$, rather than the free electron hole-density value $1-p$ \cite{Sachdev_1994,Wen_1996,Lee_1998,lee2004doping,Kaul_2007,Qi_2010,Mei_2012}. When $B$ condenses, the gauge symmetry is fully broken, and various symmetry breaking orders including $d$-wave superconductivity and charge order can be inherited by the electrons. Within this approach, superconductivity and charge order are treated on equal footing and can be viewed as low temperature, competing instabilities of a fractionalized Fermi liquid (FL$^*$) pseudogap phase. 
(Previous work \cite{Wen_1996,Lee_1998,lee2004doping} has considered the condensation of such a boson from an incoherent normal state which does not have pocket Fermi surfaces of electrons or holes, and with a
U(1) staggered flux spin liquid rather than the $\pi$-flux spin liquid. The staggered flux spin liquid has a charge $e$ boson whose condensation
leads to $d$-wave superconductivity, but not the additional possibility of charge order; moreover,
it has a trivial monopole instability \cite{song1}, so is unlikely to have significant regime of stability.) 

In this work, we will consider the transition  from the pseudogap phase with electron and/or hole pockets and a $\pi$-flux spin liquid, to a conventional $d$-wave superconductor. We will compute electronic observables in the superconducting phase via the framework of the Ancilla model \cite{Zhang_2020,Zhang_2020b,Mascot_2022,Nikolaenko_2021,Nikolaenko_2023,zhou2023ancilla,QPMbook}. While earlier work \cite{Chatterjee_2016} has considered superconductivity as a similar confinement transition from a phenomenological model of the pseudogap Fermi surfaces, the Ancilla model has the benefit of providing a microscopic model for the complete fermion dispersion in the Brillouin zone which emerges in an approximation of the Hubbard model \cite{Mascot_2022}.

The rest of this paper will be organized as follows. In Sec.~\ref{sec:AncillaModel} we will introduce the Ancilla model, and in Sec.~\ref{sec:mft} its mean-field representation which we will use to compute various electronic properties of the pseudogap normal state and $d$-wave superconductor.

In Sec.~\ref{sec:spectra}, we will describe the phenomenology of our theory on the hole-doped side, where the pseudogap normal state is captured by hole-like pockets enclosing a volume associated with hole density $p$. We will show in the framework of the Ancilla model that the hole pocket Fermi surfaces of the pseudogap undergo a transition first to a $d$-wave superconductor with 12 nodes, and then to 4 nodes as the strength of the superconducting pairing is increased. We will also compute how the Fermi velocity and $v_\Delta$ of these nodes evolves with the superconducting pairing strength.

In Sec~\ref{sec:spectrael} we will turn our focus to the electron doped side of the cuprate phase diagram. In this case, the normal state Fermi surface will be an FL$^*$ state with either ({\it i\/}) only electron-like pockets in the anti-nodal region of the Brillouin zone centered at wavevectors $(0,\pi)$ and $(\pi,0)$, or ({\it ii\/}) both anti-nodal electron-like pockets and hole-like pockets in the nodal region \cite{Shen19,Armitage_2002,PhysRevB.75.224514,PhysRevLett.118.137001,Li_2019,Kartsovnik_2011,Breznay_2019,PhysRevB.92.094501}. Perhaps surprisingly, we find that even in the first case where the normal state Fermi surface only exists at the anti-nodal region and any states in the nodal region are fully gapped, a condensed superconducting pairing will immediately lead to the re-emergence of nodes near $(\frac{\pi}{2},\frac{\pi}{2})$, while the anti-nodal region is gapped out by the pairing. We will also explore how the velocities of these nodes evolve as a function of the $B$ condensate in the electron doped case.

\section{Results}

\subsection{Ancilla Model}
\label{sec:AncillaModel}

\begin{figure}[t]
    \centering
    \includegraphics[width=\linewidth]{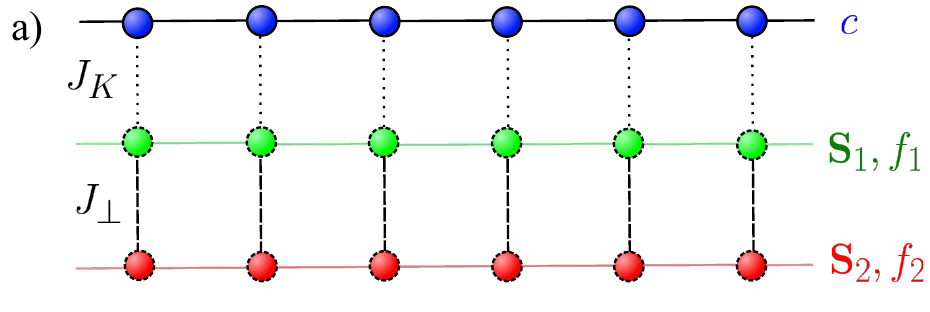}
       \caption{\textbf{Schematic of Ancilla model} | We show a schematic of the Ancilla model (a) described in Sec.~\ref{sec:AncillaModel}. We have colored the $c$ electrons blue, the first layer of spins $\vec{S}_1$ green, and the second layer of spins $\vec{S}_2$ red. The green and red bonds in the first and second layer of spins represent Heisenberg exchange interactions. The dashed line between the first and second layers of spins represent antiferromagnetic coupling $J_\perp$ and the dotted lines between the physical $c$ electrons and the first layer of spins represent Kondo coupling $J_K$. The lines connecting the $c$ electron sites denote the $c$ electrons' hopping. The red, green, blue coloring corresponding to the third, second, and first layers respectively will be kept consistent throughout the paper. The original model is a Hubbard on the $c$ layer, and the Hubbard $U$ has been canonically transformed away by adding two Ancilla layers of a bilayer antiferromagnet \cite{Nikolaenko_2021}.}
    \label{fig:Ancilla}
\end{figure}

In this section we will discuss the model we use to compute the spectral properties of the $d$-wave superconductor. The model is the Ancilla model of Ref.~\cite{Zhang_2020}, which has been shown to have all the ingredients needed to reproduce photo-emission data in both the pseudogap and Fermi liquid regime \cite{Mascot_2022,Nikolaenko_2021}. A schematic of the model is shown in Fig.~\ref{fig:Ancilla}. 

The general idea is to map the low energy physics of the single-band Hubbard-like model of the $c$ layer to a model with free electrons on the $c$ layer coupled to a bilayer square lattice antiferromagnet of the $\vec{S}_1$ and $\vec{S}_2$ layers. But we emphasize that the 
$\vec{S}_1$ and $\vec{S}_2$ layers are just {\it ancilla qubits} {\it i.e.} they are useful quantum degrees of freedom employed at intermediate stages to obtain a wavefunction with non-trivial entanglement on the $c$ layer alone. 
There is a passing similarity to `hidden layers' in current models of machine learning wavefunctions \cite{Moreno:2021jas}, but it is important that in our case the hidden layers are quantum, not classical, as that is the key to obeying the Luttinger-Oshikawa constraints on Fermi surface volumes.
An explicit example of an ancilla wavefunction for the FL* pseudogap metal was presented in Ref.~\cite{Zhang_2020}: a product of Slater determinants on the top-two and bottom layers was projected onto the physical top $c$ layer by taking the overlap with rung singlets on the $\vec{S}_1$ and $\vec{S}_2$ layers. In the analytical theories \cite{Zhang_2020,Zhang_2020b,QPMbook} of the Ancilla model, the projection is performed by emergent gauge fields. In the following, we shall not consider the influence of the emergent gauge fields as they are higgsed in all the phases considered here. Consequently, we maintain that the low-energy dispersions of the fermionic excitations described below will apply also to the single-band Hubbard model. 

An earlier work \cite{Qi_2010} obtained formally the same FL* pseudogap metal directly in the single-band model, while using the bosonic spinon $\mathbb{CP}^1$ formulation of the spin liquid. But this single-band formulation does not yield a theory for a transition from FL* to the large doping Fermi liquid, as is possible in the Ancilla model \cite{Zhang_2020,Zhang_2020b}.
Moreover, the bosonic spinon spin liquid is dual to the fermionic spinon $\pi$-flux spin liquid \cite{Wang_2017}, and the mapping of the bosonic spinon FL* formulation to the fermionic spinon theory used here becomes evident in the Ancilla model. The fermionic spinons are essential to obtaining the FL* electronic spectrum across the Brillouin zone, and not just near the zone diagonals. The fermionic spinons are also key to obtaining 4 nodal points in the $d$-wave superconductor (in contrast to the 8 nodal points in the SC* state obtained by pairing Fermi pockets in the bosonic spinon approach \cite{Moon11}).  

As has been discussed in  Ref.~\cite{Nikolaenko_2021}, we can derive the Ancilla model by an extension of the method used to introduce paramagnons in the theory correlated metals. We start from single-band Hubbard model
\begin{equation}\label{Hubbard}
    H_U =-\sum_{\vec{i},\vec{j}}\sum_{\alpha}t^c_{\vec{i},\vec{j}}c^\dagger_{\vec{i},\alpha} c_{\vec{j},\alpha} + U \sum_{\vec{i}} n_{\vec{i} \uparrow} n_{\vec{i} \downarrow}
\end{equation}
and exactly decouple the Hubbard term by the paramagnon field $\vec{\Phi}_i$
\begin{align}
    H_{\vec{\Phi}} = & -\sum_{\vec{i},\vec{j}}\sum_{\alpha}t^c_{\vec{i},\vec{j}}c^\dagger_{\vec{i},\alpha} c_{\vec{j},\alpha} \nonumber \\
    & + \sum_{\vec{i}} \left[
    \frac{3}{8U} \vec{\Phi}_i^2  - \vec{\Phi}_i \cdot c^\dagger_{\vec{i},\alpha} \frac{\vec{\sigma}^{\alpha\beta}}{2}c_{\vec{i},\beta} \right]\,. \label{Paramagnon}
\end{align}
In the traditional paramagnon method, $\vec{\Phi}$ is treated as a nearly Gaussian field whose correlators are damped by coupling to the Fermi surface of $c$. Here we identify the $\vec{\Phi}$ paramagnon with the rung-triplet excitation of the $\vec{S}_1$ and $\vec{S}_2$ layers in Fig.~\ref{fig:Ancilla}; such a triplet excitation is clearly present when $J_\perp$ is large. Indeed, the model in Fig.~\ref{fig:Ancilla} (described explicitly below) can be mapped back to the single-band Hubbard model in (\ref{Hubbard}) via a Schrieffer-Wolff transformation valid for small $J_K/J_\perp$ \cite{Nikolaenko_2021}. For other values of $J_K/J_\perp$, we need the fluctuations of emergent gauge fields to project out the ancilla layers, but as argued above, we expect such gauge fluctuations to not modify the low energy excitations in the phases considered below.

The Hamiltonian of the Ancilla model in Fig.~\ref{fig:Ancilla} is:
\begin{equation}\label{Hamiltonian}
    H=-\sum_{\vec{i},\vec{j}}\sum_{\alpha}t^c_{\vec{i},\vec{j}}c^\dagger_{\vec{i},\alpha} c_{\vec{j},\alpha}+H_{c,f_1}+H_{f_1,f_2}+H_{f_1,f_1}+H_{f_2,f_2}
\end{equation}
In the above, the two spin flavors of $c$ electrons are the physical degrees of freedom which carry the doping and are fixed to have filling:
\begin{equation}\label{constraintc}
   \frac{1}{N}\sum_{\vec{i}}\sum_{\alpha}  c^\dagger_{\vec{i},\alpha}c_{\vec{i},\alpha} =1-p
\end{equation}
Where in the above $p$ denotes the hole doping. $H_{c,f_1}$ denotes a Kondo coupling between the $c$ electrons on each site and a layer of spins $S_1$:

\begin{equation}\label{Hcf1}
\begin{split}
    H_{c,f_1}&=J_K \sum_{\vec{i}} c^\dagger_{\vec{i},\alpha} \vec{\sigma}^{\alpha\beta}c_{\vec{i},\beta}\cdot \vec{S}_{1,\vec{i}}\\&=J_K \sum_ic^\dagger_{\vec{i},\alpha} \vec{\sigma}^{\alpha\beta}c_{\vec{i},\beta}\cdot f^\dagger_{1,\vec{i},\gamma}\vec{\sigma}^{\gamma\delta}f_{1,\vec{i},\delta}
\end{split}
\end{equation}
Where we have chosen to represent the spins of the first layer $\vec{S}_{1,\vec{i}}$ with fermionic spinons $f_1$, subject to the local constraint:
\begin{equation}\label{constraintf1}
     \sum_{\alpha}  f^\dagger_{1,\vec{i},\alpha}f_{1,\vec{i},\alpha} =1
\end{equation}
The term $H_{f_1,f_2}$ describes the antiferromagnetic coupling between the layer of $\vec{S}_1$ spins and a second layer of spins labeled $\vec{S}_2$:

\begin{equation}\label{Hf1f2}
\begin{split}
    H_{f_1,f_2}&=J_\perp \sum_i\vec{S}_{1,\vec{i}}\cdot \vec{S}_{2,\vec{i}}\\&=J_\perp \sum_if^\dagger_{1,\vec{i},\alpha} \vec{\sigma}^{\alpha\beta}f_{1,\vec{i},\beta}\cdot f^\dagger_{2,\vec{i},\gamma}\vec{\sigma}^{\gamma\delta}f_{2,\vec{i},\delta}
\end{split}
\end{equation}
Where we have introduced a second set of fermionic spinons $f_2$ to represent the $\vec{S}_2$, which are subject to their own local constraint:
\begin{equation}\label{constrainthidden}
    \sum_{\alpha}  f^\dagger_{2,\vec{i},\alpha}f_{2,\vec{i},\alpha} =1
\end{equation}
The terms $H_{f_1,f_1}$ and $H_{f_2,f_2}$ describe Heisenberg exchange interactions between the spins in the first and second layers respectively. In order to reproduce the Fermi arcs seen in experiments in the pseudogap phase \cite{Proust_2019}, the model must contain deconfined fractional degrees of freedom in order to violate Luttinger's theorem \cite{Senthil_2003,Senthil_2004}. We will therefore take $H_{f_2,f_2}$ to be described by the $\pi$-flux spin liquid \cite{PhysRevB.37.3774}
\begin{align}
H_{f_2,f_2} = -it^{f_2}\sum_{\langle \vec{i},\vec{j}\rangle}f^\dagger_{2,\vec{i},\alpha}e_{\vec{i},\vec{j}}f_{2,\vec{j},\alpha}\,.
\end{align}
We have written the saddle point $\pi$-flux spin liquid in the second layer of spins in the gauge previously used in \cite{Wang_2017,Christos:2023oru} where $e_{\vec{i},\vec{j}}=-e_{\vec{j},\vec{i}}$, $e_{\vec{i},\vec{i}+\vec{x}}=1$, $e_{\vec{i},\vec{i}+\vec{y}}=(-1)^x$.

\subsection{Mean-field theory}
\label{sec:mft}

After a mean field decoupling the model can be written in the following form:
\begin{equation}\label{MFH}
\begin{split}
H=\sum_{\vec{i},\vec{j}}\sum_\alpha&\left[-t^c_{|\vec{i}-\vec{j}|}c^\dagger_{\vec{i},\alpha}c_{\vec{j},\alpha}-t^{f_1}_{|\vec{i}-\vec{j}|}f^\dagger_{1,\vec{i},\alpha}f_{1,\vec{j},\alpha}\right]\\&-it^{f_2}\sum_{\langle \vec{i},\vec{j} \rangle}\sum_\alpha f^\dagger_{2,\vec{i},\alpha}e_{\vec{i},\vec{j}}f_{2,\vec{j},\alpha}.\\&+\sum_{\vec{i}}\sum_\alpha\left[\Phi\left(c^\dagger_{\vec{i},\alpha}f_{1,\vec{i},\alpha}+f^\dagger_{1,\vec{i},\alpha}c_{\vec{i},\alpha}\right)\right.\\&\left.+iB_{1,\vec{i}}f^\dagger_{2,\vec{i},\alpha}f_{1,\vec{i},\alpha}-iB_{2,\vec{i}}\epsilon_{\alpha\beta}f_{2,\vec{i},\alpha}f_{1,\vec{i},\beta}+h.c.\right]
\end{split}
\end{equation}

In the above the chemical potentials $\mu_c$, $\mu_{f_1}$ and $\mu_{f_2}$ must be adjusted such that Eq.~\ref{constraintc}, ~\ref{constraintf1} and ~\ref{constrainthidden} are satisfied. In practice, we set the chemical potentials on an 80$\times$80 momentum space grid and set an error threshold of .01 for each filling.
We also allow for next, next-next, and next-next-next nearest neighbor terms for the $c$ and $f_1$ electron dispersions. For the hole-doped system, the hopping parameters in \cite{Mascot_2022} were found to best match photo-emission data taken in the pseudogap regime \cite{He_2011}. Therefore, for the hole-doped case, we will take $t^c_{0,1}=.22$ eV, $t^c_{1,1}=-.034$ eV, $t^c_{2,0}=.036$ eV, and $t^c_{2,1}=-.007$ eV for the $c$ electron dispersion and $t^{f_1}_{0,1}=-.1$ eV, $t^{f_1}_{1,1}=.03$ eV, and $t^{f_1}_{2,0}=.01$ eV. The above hoppings were fit to photo-emission data assuming $\Phi=.09$ and $p=.206$. We have not included the pairing terms which will appear in a mean-field decoupling of the first layer Heisenberg interactions, as there should be no pairing in the first layer in the pseudogap phase. In the second Ancilla layer, we have taken $t^{f_2}=.14$ eV. The $f_1$ and $f_2$ spinons are coupled via the two component, complex boson $B_i=\begin{pmatrix}
    B_{1,\vec{i}}\\B_{2,\vec{i}}
\end{pmatrix}$ which is spin singlet under global SU(2) spin rotations. $B_i$ is viewed as the Higgs field in the context of this theory. 

Various possible phases of Eq.~\ref{MFH} have been studied in previous work on the Ancilla model \cite{Zhang_2020,Mascot_2022,Nikolaenko_2021}. The pseudogap phase corresponds to the case where $J_K$ is much larger than $J_\perp$. In this case $B$ is gapped but $\Phi$ is condensed and the gauge symmetry of the spin liquid is unbroken. The ground state in this case is described by a fractional Fermi liquid (FL$^*$) state, and it is possible to choose parameters such that the electrons will hybridize with the $f_1$ spinons and form hole-like pockets with associated hole density $p$, where the spectral weight of the $c$ electrons is highest on the front-side pocket closest to the center of the Brillouin zone as in the first row of Fig.~\ref{fig:densities}. A Fermi liquid can be realized in the case where $J_\perp$ in Eq.~\ref{Hamiltonian} is much larger than $J_K$, which leads to $\Phi$ becoming gapped. In this case, the $f_1$ and $f_2$ spinons will form singlets at each site, and the $c$ electrons will exhibit a conventional Fermi liquid Fermi surface with hole density $1-p$.

In this work, we will consider starting from a normal state where only $\Phi$ is condensed such that the Fermi surface has hole-density $1-p$ and ask how the electronic spectrum evolves as a $d$-wave superconductor sets in when $B$ condenses on top of this normal state. As discussed in previous work \cite{Christos:2023oru} for the case of the $\pi$-flux spin liquid and in \cite{Wen_1996,Lee_1998,lee2004doping} for the case of the U(1) staggered flux spin liquid, the different ways in which the two components of $B$ condense can break different symmetries corresponding to distinct orders which may be inherited by the physical electrons if $\Phi$ is also condensed. Expanding about the two band minima of the $\pi$-flux dispersion, we have the following expression for the chargon $B$ in terms of $B_{a+}$ and $B_{a-}$, the continuum degrees of freedom associated with the $+$ and $-$ minima of the $\pi$-flux mean field dispersion:
\begin{equation}
\begin{split}
    B_{a}(\vec{r})=-B_{a,+}e^{i\pi(x+y)/2}+(1+&\sqrt{2})B_{a,-}e^{i\pi(x-y)/2}\\& \text{ for $x$ even}
\end{split}
\end{equation}
\begin{equation}
\begin{split}
    B_{a}(\vec{r})=(1+\sqrt{2})B_{a,+}e^{i\pi(x+y)/2}&-B_{a,-}e^{i\pi(x-y)/2}\\& \text{ for $x$ odd}
\end{split}
\end{equation}
In the above $a$ is a label that runs over Nambu gauge indices. We will focus on the case where $B$ condenses in such a way that a $d$-wave pairing is inherited by the physical electrons. In this case the following continuum order parameter will be condensed:
\begin{equation}
   \Delta= \epsilon_{ab}B_{a+}B_{b-}
\end{equation}
We can then choose 
\begin{align} 
B_{a+}=\frac{1}{\sqrt{2}}(-b,b)\, \quad B_{a-}=\frac{1}{\sqrt{2}}(b,b) \label{defBb}
\end{align}
as a mean-field ansatz for a pairing which will be inherited by the $c$ electrons and ask how the electronic observables will evolve when $b$ is nonzero.

\subsection{Superconductor spectra with hole-doping}\label{sec:spectra}
\begin{figure}[H]
    \centering
    \includegraphics[width=\linewidth]{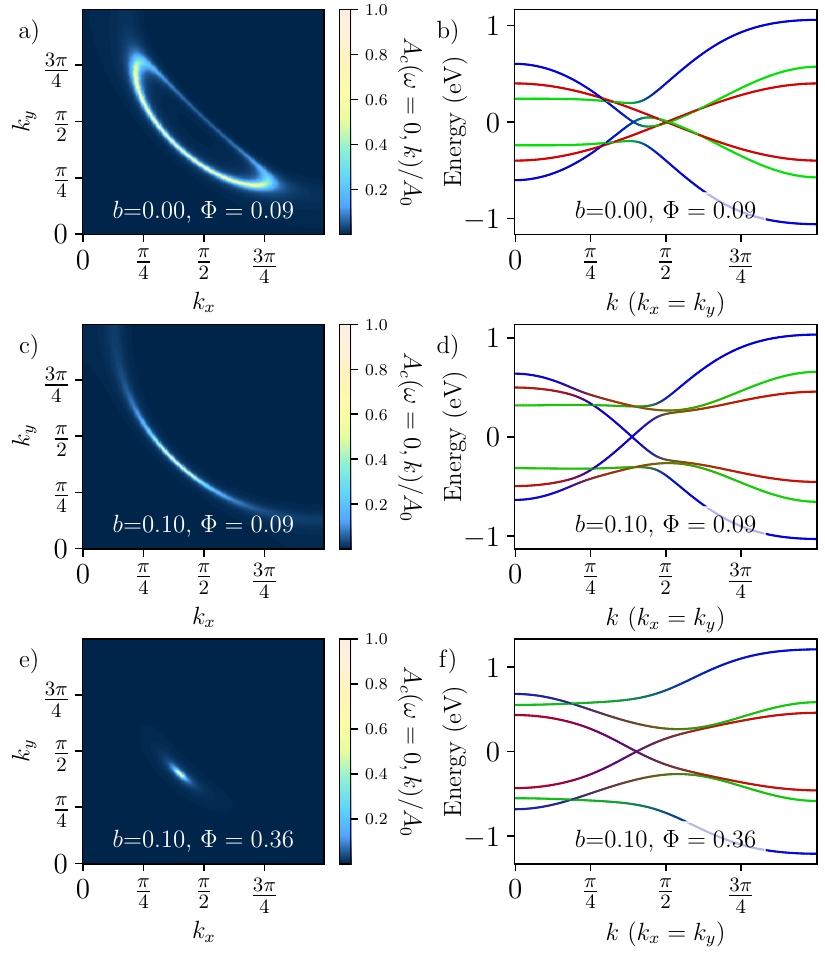}
    \caption{\textbf{Spectral density in superconductor for hole-doped normal state with hole pockets} | We show the electron spectral density (a,c,e) and band structure along a diagaonal cut in the Brilloun zone (b,d,f) for different values of $b$ and $\Phi$ (see Eq.~\ref{defBb} for the definition of $b$). In the band structure plots, the bands are colored with rgb values where blue=$c$ electrons, green=$f_1$ fermions, and red=$f_2$ fermions. We have plotted all bands in the Nambu basis, thus the spectrum is always particle-hole symmetric. When $B=0$ but $\Phi>0$ (top row), the electron spectral function shows hole-like Fermi pockets in the nodal region of the Brillouin zone formed from hybridization between the $c$ and $f_1$ electrons. When $b$ is nonzero and a $d$-wave superconducting order is inherited by the $c$ fermions (middle and bottom rows), all of the states at the Fermi level are gapped out except for a node on the front side of the original hole pocket of the parent state. The $c$ electron velocity perpendicular to the $k_x=k_y$ cut is shown to increase if $\Phi$ is made larger (bottom row). Spectral densities are normalized by their maximum value $A_0$. All spectral densities are computed with a lifetime parameter $.005i$. In practice when plotting dispersion and spectral functions, we use the gauge of \cite{Wen_2002} which is manifestly translationally invariant for the $\pi$-flux spin liquid dispersion, though we note in the case when $B$ is not condensed, the spinon bands are not gauge invariant and do not appear in any physical observable.}
    \label{fig:densities}
\end{figure}

In this section we will discuss some qualitative features of the superconductor which sets in when $B$ condenses in a normal state where $\Phi\neq0$. We show an example in Fig.~\ref{fig:densities} of the electronic spectral density over the transition of an FL$^*$ normal state to a $d$-wave superconductor as $B$ is condensed.

There are several features of the electron spectra which are of particular relevance to experiments. 
One important question we will address is the number of nodes our theory predicts will appear where $B$ is condensed, given the experimental evidence for 4 nodes \cite{Damascelli_2003,Vishik_2010} in the Brillouin zone in the hole-doped superconducting state. We will also study the evolution of the velocities $v_F$ and $v_\Delta$ as $b$ becomes nonzero as well as discuss the phenomenology of the pairing on the electron-doped side of the phase diagram.

\subsubsection{Number of nodes}
\begin{figure*}[t]
    \centering
    \includegraphics[width=\linewidth]{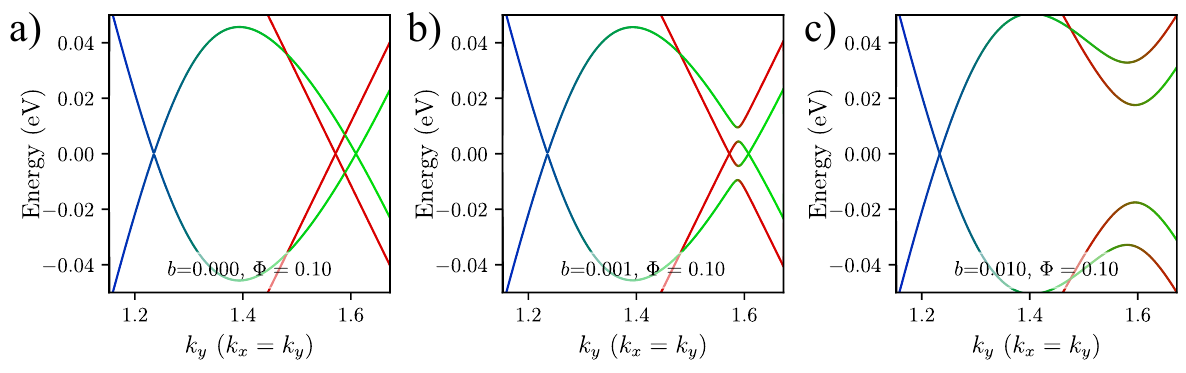}
    \caption{\textbf{Number of nodes in hole-doped superconductor} | We show the evolution of the Fermi surface along a diagonal cut through the nodal region of the Brillouin zone, focusing on the region where the hole pocket in the normal state is present. (a) shows the Fermi surface formed from the hybridized $c$ electrons and $f_1$ spinons with an isolated the Dirac cone from the $\pi$-flux spin liquid of the $f_2$ layer. For very small values of $b$, when the $f_2$ spinons hybridize with the $c$ electrons and $f_1$ spinons, the Fermi surface has 12 nodes as shown in (b). For larger values of $b$, there are 4 nodes on the Fermi surface as shown in (c). }
    \label{fig:Phismall}
\end{figure*}

The first question we will address is how many nodes there are in the superconducting state. We find that similar to past studies \cite{Chatterjee_2016}, the answer to this question depends on the values of $b$ and $\Phi$. There are two possibilities for our chosen parameters which are depicted in Fig.~\ref{fig:Phismall}.

If the particle-hole symmetry breaking in the first layer of spinons is taken to be small, then there is a small but finite window of $0<b<b_c$ where the spectra shows 3 nodes in each quadrant of the Brillouin zone, or 12 nodes total. However, the appearance of a window of $b$ with 12 nodes results from the particle-hole asymmetry of the $f_1$ spinon bands which was found to be small when next and next-next nearest neighbor hoppings in the second layer were fit to experiment \cite{Mascot_2022}. Thus this feature persists for a very small window of $b$ before two of the three nodes in each quadrant of the Brillouin zone annihilate and we are left with a spectrum with 4 nodes, the scenario which is born out in experiments \cite{Damascelli_2003,Vishik_2010}. 

In the above discussion on number of nodes, we have assumed $t^{f_2}=.14$ eV. While the overall magnitude of $t^{f_2}$ should not change the number of nodes, the sign of $t^{f_2}$ will determine which 2 of the 3 nodes along the diagonal annihilate first and therefore qualitatively change the mean-field dispersion. Since the case where $t^{f_2}<0$ seems not to display the universal behavior discussed above, we consider it separately in Supplement \cite{supplement1}. Additional plots of the dispersion throughout the Brillouin zone can also be found in \cite{supplement1}.

\subsubsection{Velocities of node}
\begin{figure}[tb]
    \centering
    \includegraphics[width=\linewidth]{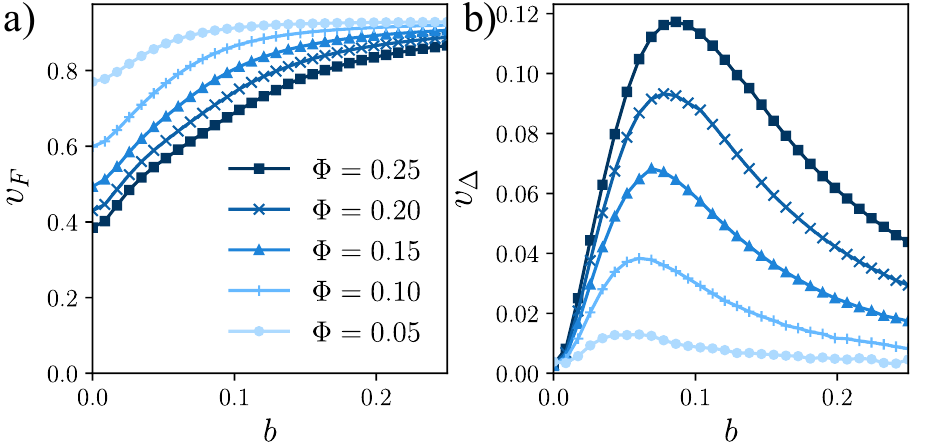}
    \caption{\textbf{Node velocities in superconductor for hole-doped normal state} | We show the velocities $v_F$ (a) and $v_\Delta$ (b) as a function of $b$ for different values of $\Phi$. For small $b$, $v_F$ takes on the Fermi velocity of the normal state Fermi surface at $k_x=k_y$, while $v_\Delta$ begins at 0 when $B=0$ and increases with finite $b$ as the effective pairing grows. For large $b$, $v_F$ approaches the Fermi velocity of the un-hybridized $c$ electron bands, while $v_\Delta$ tends to 0. }
    \label{fig:Velocities}
\end{figure}

In this section we will discuss the evolution of the Fermi velocities as $b$ is increased for various values of $\Phi$. Our results are shown in Fig.~\ref{fig:Velocities}.

There will be two independent velocities, $v_F$ which is defined as the velocity parallel to the $k_x=k_y$ contour in the Brillouin zone and $v_\Delta$, the velocity perpendicular to this contour. Since a region of 12 nodes appears for a relatively small window of $b$, we consider the velocity only in the case of the node which falls along the original $c$ electron Fermi surface and has the highest overlap with the $c$ electrons. We compute $v_F$ and $v_\Delta$ by discretizing diagonal cuts through a quadrant in the Brillouin zone into 80,000 momentum points and performing a least squares fit on the 500 momentum points nearest the node. 

The nodal velocity perpendicular to $k_x=k_y$, $v_\Delta$, begins at zero when $b=0$. When $B$ is condensed, and $b$ is small relative to $\Phi$, the superconducting pairing is inherited by the $c$ electrons and as a result $v_\Delta$ becomes finite and increases with $b$ as the effective pairing gaps out any states which are not on the Brillouin zone diagonal. $v_\Delta$ will continue to increase until $b$ is roughly of the same order as $\Phi$ where $v_\Delta$ attains a maximum. When $b$ is sufficiently large relative to $\Phi$, $v_\Delta$ will begin to decrease as the layer of $c$ electrons becomes effectively decoupled from the first and second layer of spinons which are pushed away from the Fermi level. For large enough $b$, the $c$ electrons spectral density will resemble the original Fermi surface of the decoupled $c$ electrons and $v_\Delta$ will tend towards zero as $b$ increases.

The nodal velocity along $k_x=k_y$, $v_F$, begins at a finite value defined by the normal state Fermi velocity and monotonically increases with $b$ until it saturates in the limit where $b\gg\Phi$ to the value of the Fermi velocity of the decoupled $c$ electron bands at the Fermi surface. The ratio of $v_\Delta$ to $v_F$ is small for all values of $b$ as the Fermi velocity $v_F$ originates mostly from the Fermi velocity of the $c$ electrons while the velocity $v_\Delta$ is 0 in the normal state and is a higher order effect in $\Phi$ and $b$.

\begin{figure*}[tb]
    \centering
    \includegraphics[width=\linewidth]{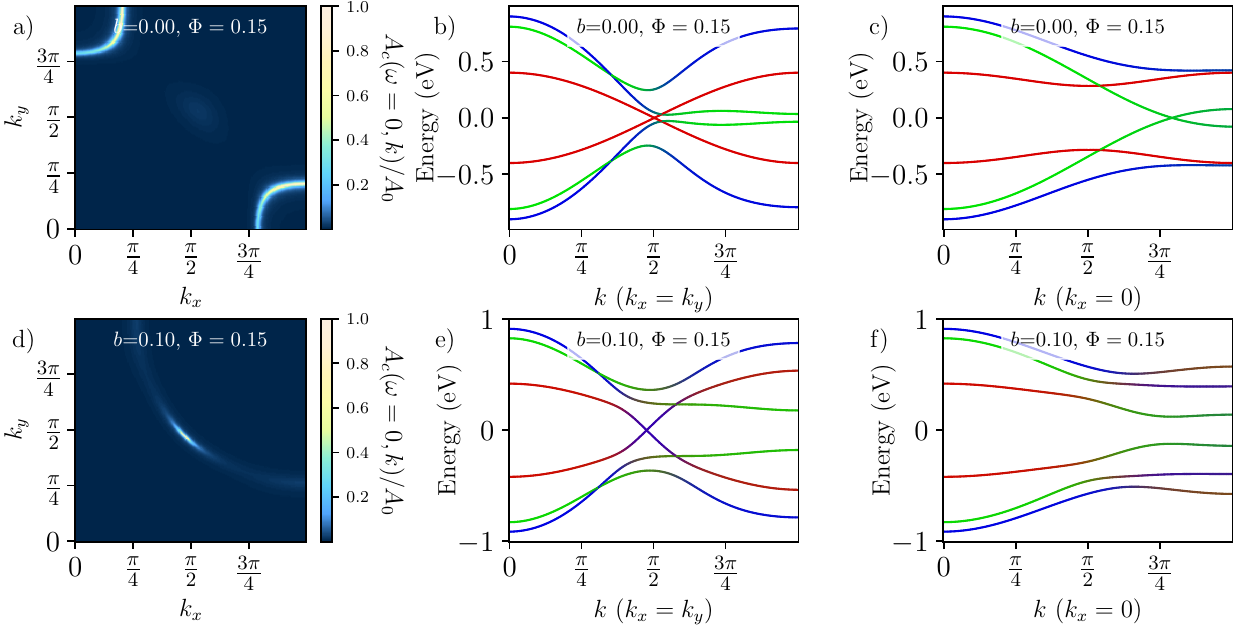}
    \caption{\textbf{Spectral density in superconductor for electron-doped normal state with only electron pockets} | Evolution of spectral density of $c$ electrons (a,d), dispersion along a diagonal cut through the Brillouin zone (b,e), and dispersion along a vertical cut through the $(0,\pi)$ (c,f) for $B=0$ (top row) and $B>0$ (bottom row). All plots are computed at \textit{electron doping} $p=.15$. We note the nodes which for finite $b$ along the diagonal which were not previously present in the $c$ electron density in the normal state.}
    \label{fig:densitieselectron}
\end{figure*}

\subsection{Superconductor spectra with electron-doping}\label{sec:spectrael}
\begin{figure*}
    \centering
    \includegraphics[width=\linewidth]{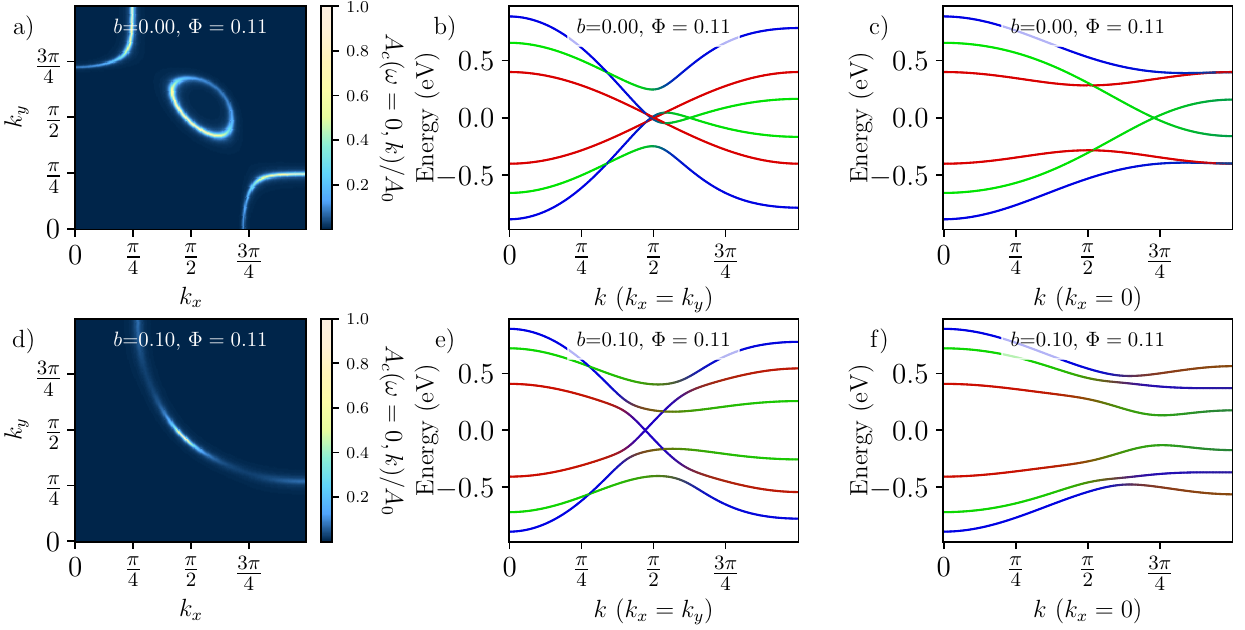}
    \caption{\textbf{Spectral density in superconductor for electron-doped normal state with both electron and hole pockets} | We show the evolution of the electron spectral density (a,d) and dispersion when superconductivity sets in for the case of a normal state as \textit{positive} electron doping which has both a hole-like pocket in the nodal region and an electron-like pocket in the anti-nodal region at electron doping $p=.15$. Dispersions are shown for a cut along the diagonal of the Brillouin zone (b,e) and for a cut which connects the anti-node and Brillouin zone center (c,f). We note that while the second row electronic spectral density shows a finite electronic spectral weight in other regions of the Brillouin zone than the node, all other points except the node have a finite, albeit sometimes small gap.}
    \label{fig:twoFS}
\end{figure*}

In this section we will discuss the spectra of an FL$^*$ to SC transition on the electron-doped side of the phase diagram. The principal difference from the hole-doped case is that we now expect instead of having hole-like pockets near $(\frac{\pi}{2},\frac{\pi}{2})$, we will have either electron pockets in the anti-nodal region of the Brillouin zone near $(0,\pi)$ and $(\pi,0)$ as in the first row of Fig.~\ref{fig:densitieselectron} or both electron like pockets at the anti-node and hole-like pockets in the nodal region as in the first row of Fig.~\ref{fig:twoFS} \cite{Armitage_2002,PhysRevB.75.224514,PhysRevLett.118.137001,Li_2019,Kartsovnik_2011,Breznay_2019,PhysRevB.92.094501}. 

For our computations with a normal state with only electron pockets, we will take the same $c$ electron hoppings as on the hole-doped side but change the $f_1$ spinon hoppings, since these were previously obtained by fitting photoemission data taken on hole-doped cuprates \cite{Mascot_2022}, and have no established values on the electron-doped side. In the first layer of spinons $t^{f_1}_{1,0}=-.1$ eV as before, but choose next nearest neighbor hopping $t^{f_1}_{1,1}=-.07$ eV and set all other hoppings in the second layer to zero. We keep the spin liquid dispersion hopping $t^{f_2}=.14$ eV. We find whether the normal state has only electron pockets at the anti-node or both electron pockets at the anti-node and hole pockets near $(\frac{\pi}{2},\frac{\pi}{2})$ depends on the value of $\Phi$, with larger $\Phi$ gapping out the hole pockets. For our computations of a normal state with both electron and hole pockets like that of Fig.~\ref{fig:twoFS}, we take $t^{f_1}_{1,1}=-.06$ eV and $t^{f_1}_{0,2}=.02$ eV and all other parameters the same as above.

\subsubsection{Number of nodes}
We will discuss the number of nodes separately for the two types of Fermi surfaces mentioned above, beginning first with the normal state where the only Fermi surfaces are electron pockets at the anti-node.
Naively, it might be expected for this case that the $d$-wave superconductor which will be inherited by the $c$ electrons when $B$ condenses will be fully gapped; however, this is not what we observe. For any finite $b$, the electron pockets of the normal state which appeared in the anti-nodal region will become fully gapped as shown in the rightmost column of Fig.~\ref{fig:densitieselectron}, but nodes will re-appear along the diagonal in the nodal region of the Brillouin zone as shown in the central column of Fig.~\ref{fig:densitieselectron}. These nodes which at $b=0$ were associated with the Dirac points of the $\pi$-flux spin liquid will hybridize with the $c$ and $f_1$ bands but cannot be gapped unless an additional symmetry such as spin rotation symmetry is strongly broken. If the above scenario is excluded, there will always be 4 nodes on the diagonal when $B$ is condensed for a normal state with only electron pockets, assuming a positive spin liquid hopping in the gauge we have chosen. We note that for small $b$, the normal state Fermi surfaces at the anti-node have a gap which may be very small and the node which appears for small $b$ initially has a low $c$ electron spectral weight. 

For the case of a normal state which has both electron pockets at the anti-node and hole pockets at the node, we observe the same transition from 4 to 12 nodes as $b$ is increased as we observed in the hole-doped case.

In all of our analysis on the electron-doped side of the phase diagram, we have assumed $t^{f_2}=.14$. However, 
changing the sign of $t^{f_2}$ will result in qualitatively different behavior in the number of nodes similar to the hole-doped case as shown in the Supplement \cite{supplement1}. However, as was shown in Appendix 3 of \cite{Christos:2023oru}, only the former sign of $t^{f_2}$ corresponds to a chargon potential which favors the continuum superconductor ansatz we have taken here.

\begin{figure}[tb]
    \centering
    \includegraphics[width=\linewidth]{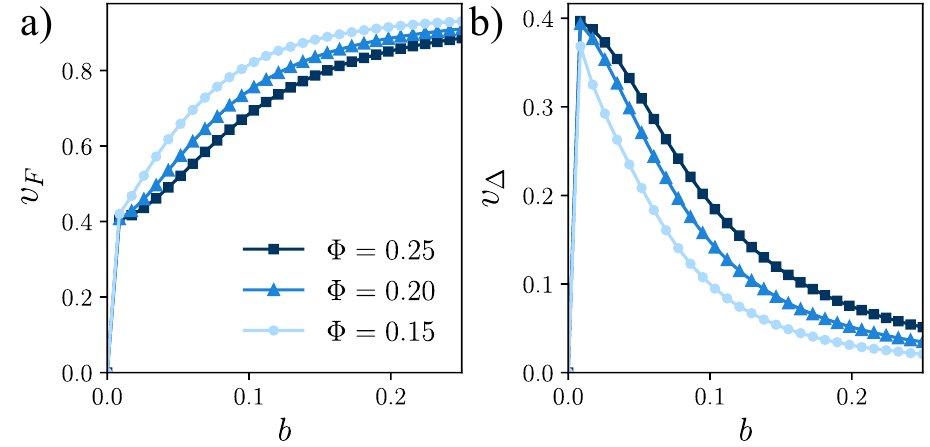}
    \caption{\textbf{Node velocities in superconductor for electron-doped normal state} | We show $v_F$ (a) and $v_\Delta$ (b) for positive electron doping as a function of $b$ for several different values of $\Phi$. In the limit of large $b$, both velocities show the same asymptotic behavior as in the hole-doped case.}
    \label{fig:velocitieselectron}
\end{figure}

\subsubsection{Velocities of node}
We also show how $v_F$ and $v_\Delta$ of the superconductor nodes on the Brillouin zone diagonal evolve for positive electron doping as a function of $b$ and for different values of $\Phi$ in Fig.~\ref{fig:velocitieselectron} for the choice of normal state with only electron pockets. For this choice of normal state, we study a narrower range of $\Phi$ than in the hole-doped case, since we wish to choose $\Phi$ such that the normal state is gapped at the node. The velocities in the case for which there are additional hole pockets is similar to the behavior shown in Fig.~\ref{fig:Velocities}. Since there is no Fermi surface observable in the electron spectral density in the nodal region at $b=0$, both $v_F$ and $v_\Delta$ immediately jump to a finite value for finite $b$. For small $b$, $v_F$ and $v_\Delta$ are roughly equal as they are essentially just inherited from the $\pi$-flux spin liquid's Dirac points which are isotropic. As $b$ increases, we see $v_F$ will first slowly increase as band repulsion which flattens the velocity competes with $b$, but in the limit $b\gg\Phi$, $v_F$ ultimately returns to the Fermi velocity of the decoupled $c$ electrons as in the hole doped case. Similar to the behavior of $v_\Delta$ at large $b$ in the hole-doped case, here $v_\Delta$ decreases as $b$ increases, ultimately tending towards zero when $b\gg\Phi$.

\subsection{Excitation energy and quasi-particle residue}
\begin{figure}
    \centering
    \includegraphics[width=\linewidth]{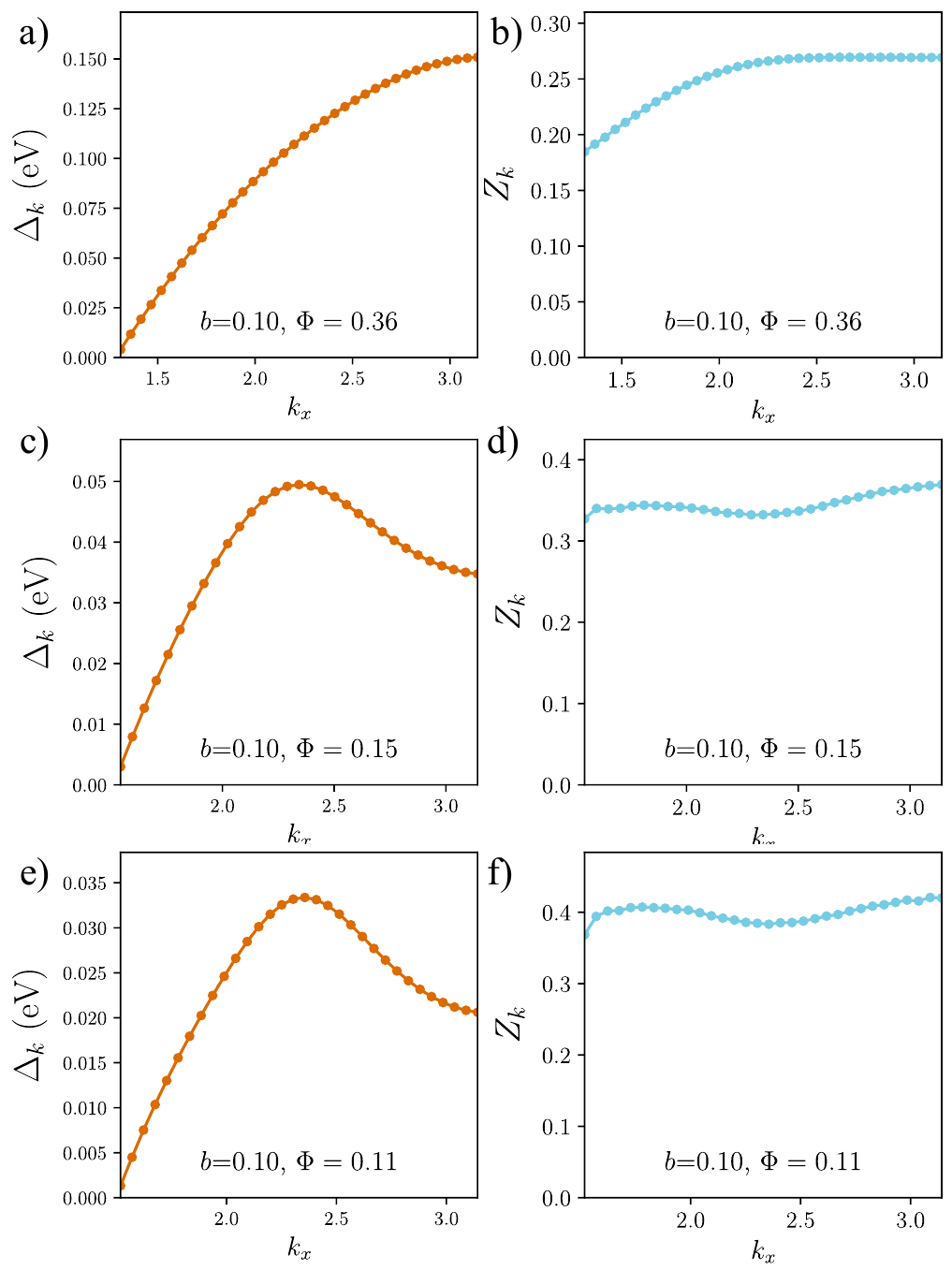}
    \caption{\textbf{Quasi-particle excitation and residue} | We show the energy of the nearest quasi-particle excitation to the Fermi level (a) and quasi-particle residue at this excitation energy (b) for the hole-doped case using the normal state shown in Fig.~\ref{fig:densities}. We also show the excitation energy (c) and quasi-particle residue (d) for the electron doped case where there is only an electron pocket at the anti-node using the normal state shown in Fig.~\ref{fig:densitieselectron}. Finally, we also show the excitation energy (e) and quasi-particle residue (f) for the normal state of Fig.~\ref{fig:twoFS} where there are both electron and hole pockets. }
    \label{fig:quasi}
\end{figure}

We also show the excitation energy and quasi-particle residue of the energetically lowest-lying excitation in the superconducting state, as shown in Fig.~\ref{fig:quasi}. We plot both quantities along a contour defined by finding the $k_x$ momentum corresponding to where the energy of the lowest lying excitation is smallest for a given $k_y$ momentum value. Effectively, this contour is well approximated by choosing momenta along the original, decoupled $c$-electron Fermi surface.  We choose the first $k_y$ value plotted to be at the location of the single node which appears in the superconducting state in either the hole-doped or electron-doped case. We may then contrast the behavior of these quantities in the hole-doped and electron-doped cases. While in the electron-doped case, the excitation energy along the above specified contour increases monotonically as momentum is varied from the location of the node at $(\frac{\pi}{2},\frac{\pi}{2})$ to the Brillouin zone edge, the electron-doped case shows non-monotonic behavior as one moves along the original $c$-electron Fermi surface, away from the node. In the electron-doped case, we observe a peak in the excitation energy roughly midway between the node and edge of the Brillouin zone, consistent with the behavior observed in \cite{Shen_2023}. In the electron-doped case the quasi-particle weight $Z_{\vec{k}}$ is mostly flat with a very slight dip between the node and anti-node, whereas in the hole-doped case, the quasi-particle weight shows monotonic behavior as momentum is varied away from the node at $(\frac{\pi}{2},\frac{\pi}{2})$.

\section{Discussion}
In this work, we have studied how various electronic observables evolve when the pseudogap metal transitions to a $d$-wave superconductor, in the framework of the Ancilla model \cite{Zhang_2020}. 

For a hole-doped normal state and positive spin liquid hopping in our chosen gauge, as discussed in Sec.~\ref{sec:spectra}, we initially find a $d$-wave superconductor with 12 nodes, and then a transition to 4 nodes as the pairing strength is increased. When the normal state is chosen to reproduce experimental photo-emission data, the regime of 12 nodes is small, and the generic case for large $b$ is a superconducting state with 4 nodes. We also found the velocities $v_F$ and $v_\Delta$ associated with the surviving nodes differ in scale, with $v_\Delta$ much smaller than $v_F$ for all values of $\Phi$ and $b$ we have studied, and tending towards 0 in the limit where $b\gg\Phi$. It is therefore clear that the velocities are not directly related to the spinon velocities in the $\pi$-flux phase, which are isotropic, and this is an important difference from earlier work \cite{Wen_1996,Lee_1998,lee2004doping}. 

We have also separately studied the FL$^*$ to superconductor transition for the electron-doped case in Sec.~\ref{sec:spectrael}. In this case, we find that a normal state with both electron-like and hole-like pockets leads to the same transition from 12 to 4 nodes as observed in the hole-doped case as a function of $b$. However, surprisingly we find that in the case with only electron-like pockets near the Brillouin zone edge, the FL$^*$ state immediately transitions to a state with 4 nodes along $k_x=k_y$, as shown in Fig.~\ref{fig:densitieselectron}. This feature, unique to the electron-doped side of the phase diagram, is striking in that nodes which are not observable in the electron spectral density in the FL$^*$ case immediately reappear for any finite $b$. Unlike in the hole-doped case, $v_F$ and $v_\Delta$ begin with nearly equal values, since the surviving node for small $b$ is associated with the spin liquid Dirac point rather than the $c$ electron Fermi surface. This aspect of the electron-doped pairing follows as a direct consequence from the mean-field dispersion of the $\pi$-flux spin liquid (though this behavior would be the same had we considered another Dirac spin liquid with the same number of nodes such as the U(1) staggered flux spin liquid state). Thus for electron-doped superconductors with only electron pockets in the normal state, it is reasonable to state that nodal Bogoliubov quasiparticles are remnants of Dirac spinons made visible by the onset of superconductivity.

All our analysis was carried out for a pseudgogap metal without without long-range antiferromagnetic order. However, the appearance of antiferromagnetic order at low temperatures within the superconducting phase (as is the case in the electron-doped cuprates) should not invalidate any of our computations, and so we believe our results should continue to apply. 

Recent numerical studies \cite{Jiang_2021,jiang2023superconducting} of an electron-doped $t$-$J$ model found robust $d$-wave supercondutivity, and the authors speculated that their $d$-wave superconductor was fully gapped. From our analysis here, we maintain that a conventional $d$-wave superconductor, with pairing strength not as large as the Fermi energy, always has 4 nodal points along the zone diagonals.
A fully gapped $d$-wave superconductor requires some additional features, and can be reached via the following routes:\\
({\it i}) We start from an FL* metal with electron pockets, and then pair the electron pockets. At this point, the $f_2$ Ancilla spin liquid is still `alive' in the superconducting state, and so such a fully gapped $d$-wave superconductor is a SC* state.
Furthermore, the $\pi$-flux spin liquid is ultimately unstable \cite{YinChen23}, and this implies that a $\pi$-flux-SC* state is not stable.\\
({\it ii}) Starting from a conventional $d$-wave superconductor with 4 nodal points, the onset of strong, co-existing antiferromagnetic order can gap out the nodes when they annihilate in pairs across the magnetic Brillouin zone boundary.\\
({\it iii}) Finally, if the pairing interaction becomes as large as the Fermi energy in a conventional $d$-wave superconductor, the four nodal points can meet at the origin (or at $(\pi, \pi)$) and annihilate with each other.\\
It appears unlikely to us that any of these 3 routes apply to the study in Refs.~\cite{Jiang_2021,jiang2023superconducting}, and so we believe their superconductor does have 4 nodal points at low temperatures, and possibly only electron pockets in the normal state.

In summary, our work has provided testable predictions of what signatures conventional $d$-wave superconductivity will have if it originates from a pseudogap phase containing fractional degrees of freedom described by the $\pi$-flux spin liquid. It would be interesting to extend the approach taken in this work to capture other relevant phases in the under-doped cuprates, such as charge order \cite{doi:10.1146/annurev-conmatphys-031115-011401}. We also note that for the case of superconductivity, the quantities we computed will not necessarily be different among different Dirac spin liquids \cite{Shackleton:2021fdh}. It is therefore interesting to conceive of experimental tests which may be capable of distinguishing between different FL$^*$ normal states. We leave these possibilities to future work.
\section{Methods}
All plots are computed from a tight-binding implementation of the model in Eq.~\ref{Hamiltonian} in momentum space using the parameters for hoppings and dopings mentioned throughout the text. Chemical potentials in each layer of the Hamiltonian described in Eq.~\ref{Hamiltonian} are determined by the doping for each set of parameters within an error threshold of .01 via the bisection method. An 80$\times$80 grid in momentum space is used to fix the chemical potentials. Spectral densities are computed as:
\begin{equation}
    A(\omega,\vec{k})=-\frac{1}{\pi}\text{Im}\left[G_{cc}(\omega,\vec{k})\right]
\end{equation}
All spectral functions are computed on a quarter of the Brillouin zone with a 200$\times$200 grid. A lifetime parameter of .005 eV is used when computing spectral densities.
The quasiparticle weight $Z_{\vec{k}}$ is computed by first computing the inverse of the greens function $G^{-1}(\omega,\vec{k})=\omega-H(\vec{k})$ which is then diagonalized by a unitary transformation $U_{\vec{k}}$ such that $U_{\vec{k}}^\dagger H(\vec{k}) U_{\vec{k}}=G^{-1}_D(\omega,
\vec{k})$ where $G^{-1}_D(\omega,
\vec{k})$ is diagonal. For an excitation at energy $\Delta_{\vec{k}}$, we the compute the quasiparticle residue as:
\begin{equation}
    Z_{\vec{k}}=U_{c\uparrow,\alpha}U^\dagger_{\alpha,c\uparrow}
\end{equation}
where $\alpha$ labels the eigenvector of $U_{\vec{k}}$ corresponding to energy $E_{\vec{k}}$.
\subsection*{Data Availability}
All data generated or analyzed during this study will be made available before publication.
\subsection*{Acknowledgements}

We thank Patrick Lee for probing questions on previous work \cite{Christos:2023oru}, Z.-X. Shen for informing us about on-going photoemission experiments in the electron-doped cuprates \cite{Shen_2023}, Steve Kivelson for discussions on Refs.~\cite{Jiang_2021,jiang2023superconducting}, and our collaborators Zhu-Xi Luo, Henry Shackleton, Ya-Hui Zhang, and Mathias Scheurer on Ref.~\cite{Christos:2023oru}. We also thank Chenyuan Li and Ilya Esterlis for helpful discussions on a related work, and Alex Nikolaenko for helpful discussions on \cite{Mascot_2022}.
This research was supported by the U.S. National Science Foundation grant No. DMR-2245246, by the Gordon and Betty Moore Foundation’s EPiQS Initiative Grant GBMF8683, and by the Simons Collaboration on Ultra-Quantum Matter which is a grant from the Simons Foundation (651440, S.S.).

\subsection*{Competing Interests Statement}
The authors declare no competing interests.
\subsection*{Author Contributions}
Both authors formulated the research and wrote the paper. M.C. performed the numerical computations.

\bibliography{ref}

\begin{appendix}
    \onecolumngrid

\section{Additional dispersion plots}
In this appendix we will provide additional plots of the dispersion of the band nearest to the Fermi level for the three types of normal states considered in the main text. In supplementary figure~\ref{fig:dispersionholedope} we show dispersion for different values of $b$ and $\Phi$ for the hole-doped normal state with hole-like pockets at $(\frac{\pi}{2},\frac{\pi}{2})$. In supplementary figure~\ref{fig:dispersionholedope} we show dispersion for different values of $b$ and $\Phi$ for the electron-doped normal state with electron-like pockets at $(0,\pi)$ and $(\pi,0)$.In supplementary figure~\ref{fig:dispersionelectrondope2} we show dispersion for different values of $b$ and $\Phi$ for the electron-doped normal state with electron-like pockets at $(0,\pi)$ and $(\pi,0)$ \textit{and} hole-like pockets at $(\frac{\pi}{2},\frac{\pi}{2})$. 

\begin{figure}[h]
    \centering
    \includegraphics[width=\linewidth]{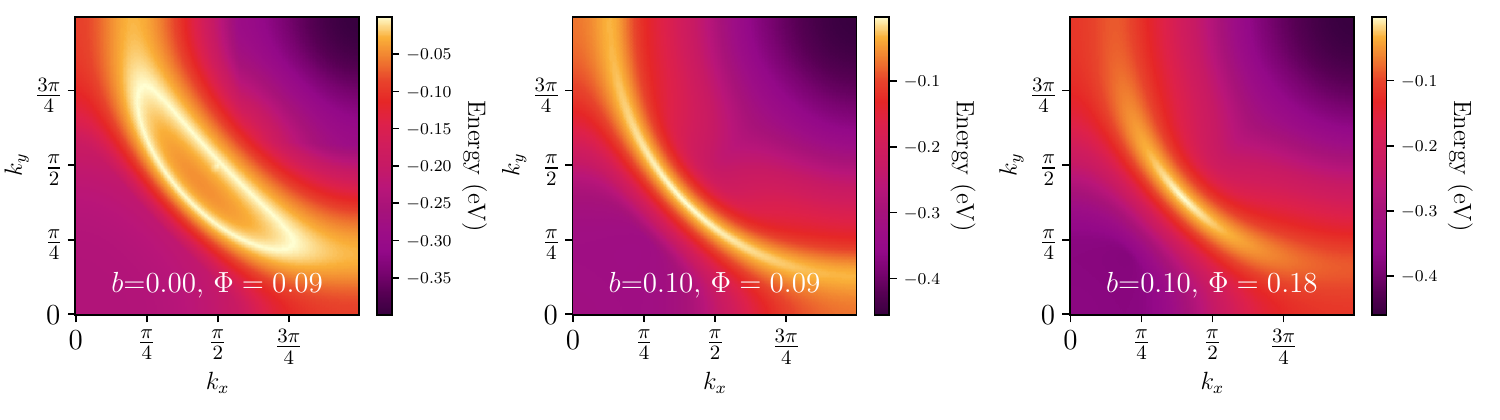}
    \caption{We plot the dispersion of the band closest to the Fermi level for different values of $b$ and $\Phi$ for hole doping $p=.2$}
    \label{fig:dispersionholedope}
\end{figure}
\begin{figure}[h]
    \centering
    \includegraphics[width=\linewidth]{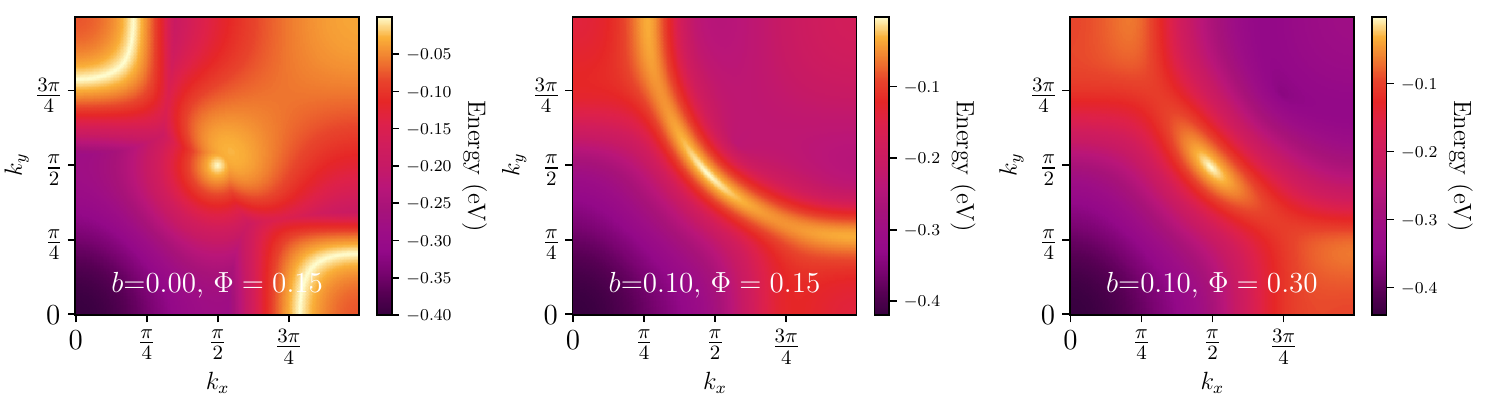}
    \caption{We plot the dispersion of the band closest to the Fermi level for different values of $b$ and $\Phi$ for electron doping $p=.15$ with a normal state with only electron pockets at the anti-node.}
    \label{fig:dispersionelectrondope}
\end{figure}
\begin{figure}[h]
    \centering
    \includegraphics[width=\linewidth]{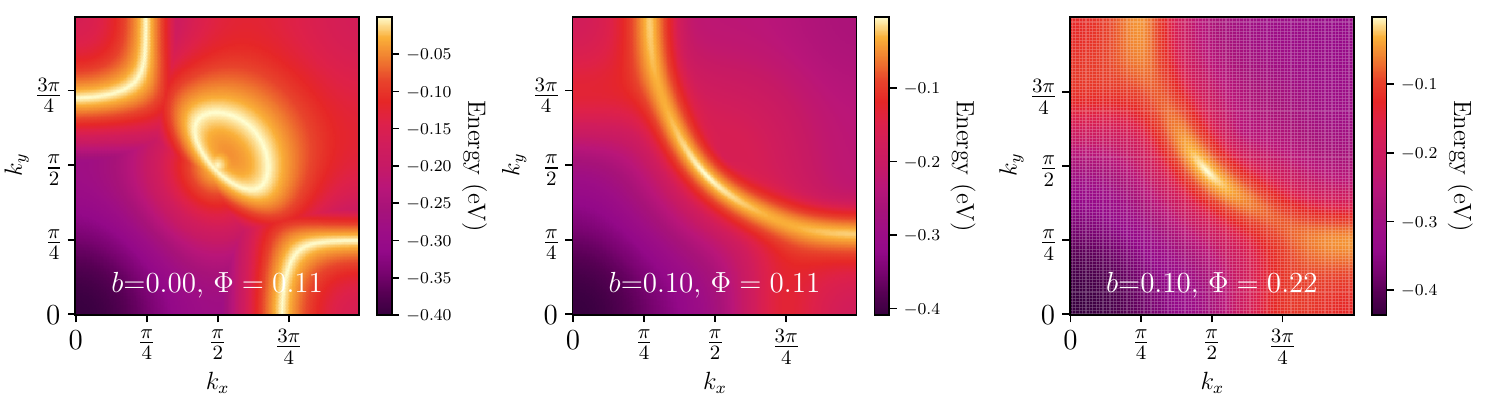}
    \caption{We plot the dispersion of the band closest to the Fermi level for different values of $b$ and $\Phi$ for electron doping $p=.15$ for a normal state with both electron and hole pockets.}
    \label{fig:dispersionelectrondope2}
\end{figure}
\section{Sign of spin liquid hopping}\label{app:oppositesign}
Throughout the main text, we have assumed $t^{f_2}$>0 for the spin liquid nearest neighbor hopping. In this appendix we will discuss how our results change if we had assumed an opposite sign of spin liquid hopping. Multiplying $t^{f_2}$ by an overall minus sign will result in the spin liquid bands changing sign. In this case case which 2 of the three nodes in the Brillouin zone are gapped out first differs from the case in the main text. In the main text, where $t^{f_2}$>0, particularly in the hole-doped case, the node in each quadrant of the Brillouin zone which survives at finite $b$ has appeared in the region of the Brillouin zone where the spectral weight of the $c$ electrons is highest. When the sign of the spin liquid hopping is flipped, the node instead first appears where the spin liquid dispersion has the highest spectral weight as shown in supplementary figure\ref{fig:oppositesign}.
\begin{figure}[tb]
    \centering
    \includegraphics[width=\linewidth]{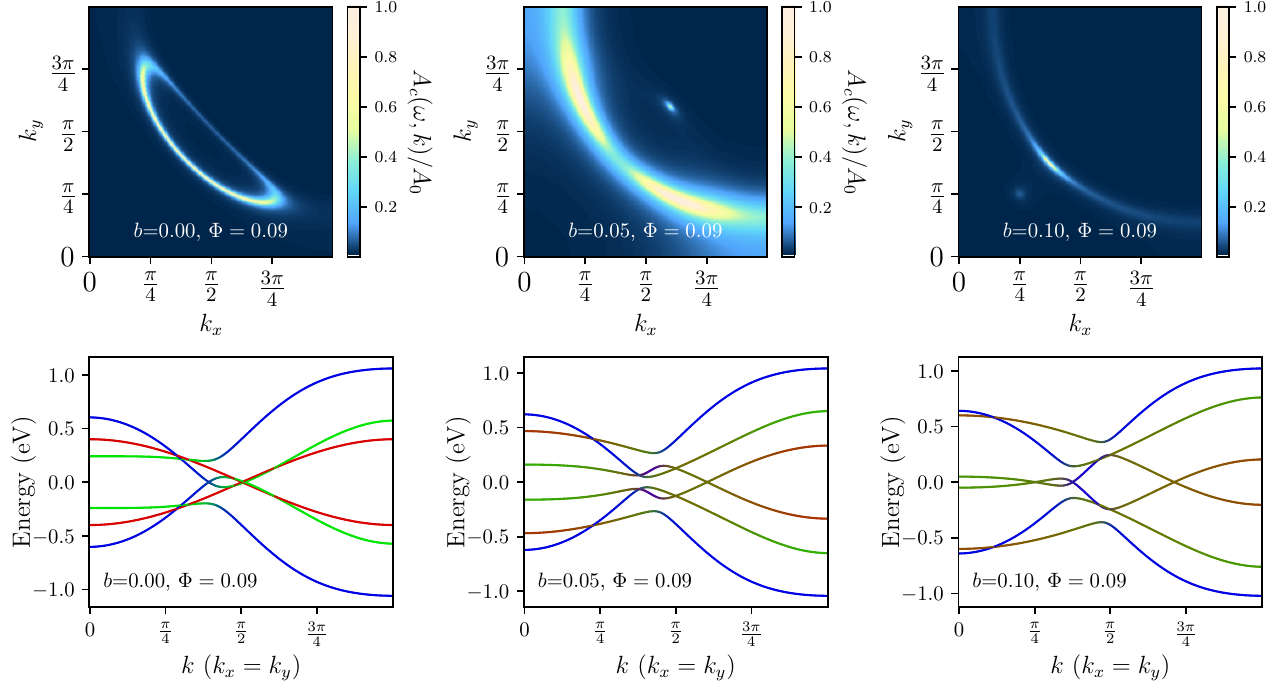}
    \caption{We show the electronic spectral density and dispersion as $b$ is increased in the case where $t^{f_2}=-.14$ for a hole-doped state. Unlike in the main text hole-doped case, the node at large $b$ appears on the backside of the normal state hole-pocket.}
    \label{fig:oppositesign}
\end{figure}

\begin{figure}[tb]
    \centering
    \includegraphics[width=\linewidth]{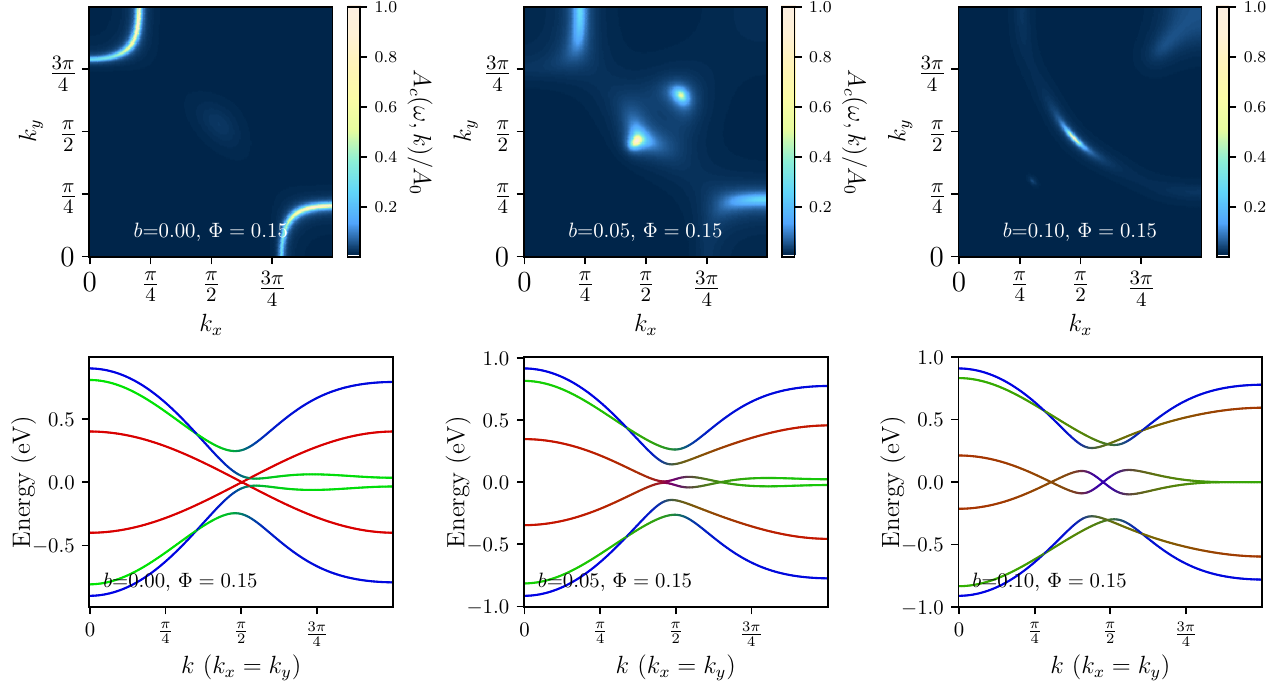}
    \caption{We show the electronic spectral density and dispersion as $b$ is increased in the case where $t^{f_2}=-.14$ for a electron-doped state with only electron-like pockets. Unlike in the main text hole-doped case, the node at large $b$ appears on the backside of the normal state hole-pocket.}
    \label{fig:oppositesignel}
\end{figure}
\begin{figure}[tb]
    \centering
    \includegraphics[width=\linewidth]{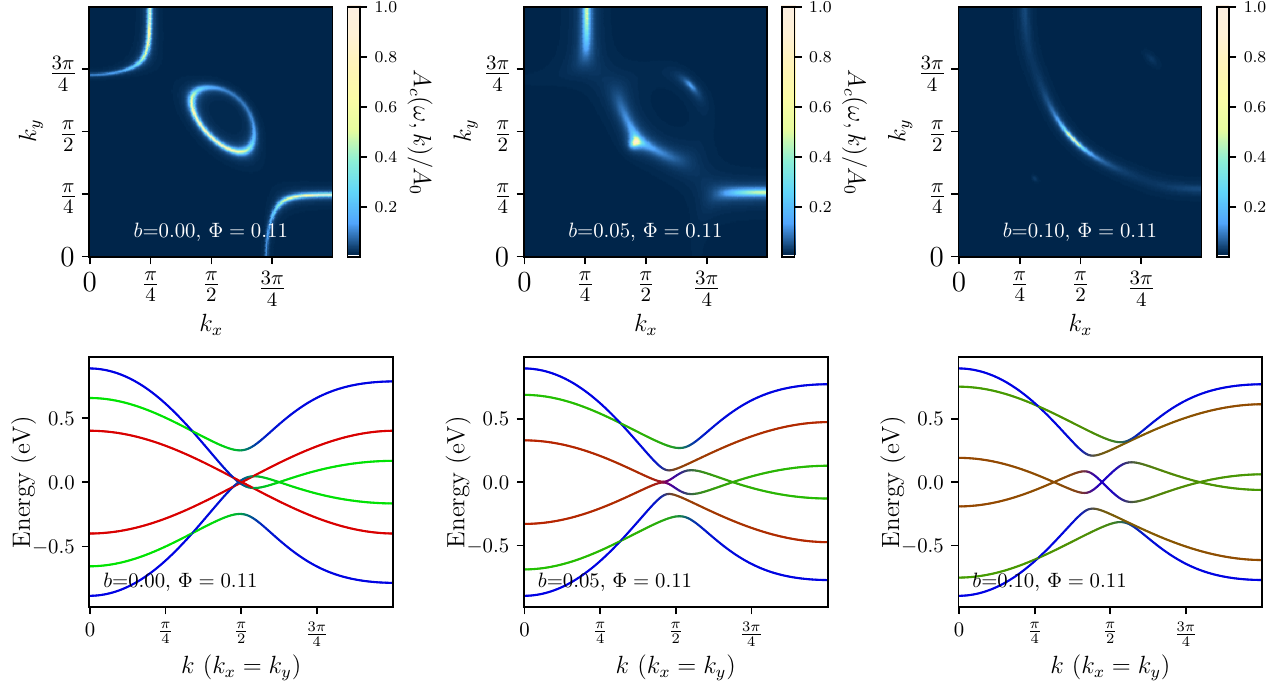}
    \caption{We show the electronic spectral density and dispersion as $b$ is increased in the case where $t^{f_2}=-.14$ for a electron-doped state with both electron-like and hole-like pockets. Unlike in the main text hole-doped case, the node at large $b$ appears on the backside of the normal state hole-pocket.}
    \label{fig:oppositesignel2}
\end{figure}
We also see that unlike in the main text case, there is a reemergence of additional allowed nodes as $b$ is further increased and $\Phi$ is kept a constant. While re-emergence of additional nodes in the main text is not strictly forbidden, we do not observe such phenomena for our choice of normal state with Fermi surfaces close to experiments. 

The electron-doped case shows different behavior with the opposite sign spin liquid hopping as well as in supplementary figure\ref{fig:oppositesignel} and supplementary figure\ref{fig:oppositesignel2}. In both cases, there are also additional nodes which appear with finite $b$ and do not seem to disappear as $b$ increases. However, the additional nodes have low spectral weight on the $c$ electrons so may not be observable experimentally.

The correct sign of $t^{f_2}$ is determined by the sign of the quadratic coupling $w_1$ in \cite{Christos:2023oru} in the chargon free energy which originates from integrating out the $c$ electrons and two layers of spinons; in particular, $w_1$ must be positive for the continuum ansatz we have used in the main text to apply. The one-loop approximation of this computation was carried out for the hole-doped spinon and electron hoppings in Appendix 3 of \cite{Christos:2023oru}, which found that assuming $t^{f_2}$>0 and led to $w_1>0$, meaning the sign used in the main text, at least for the hole-doped case is correct to some approximation. 
\end{appendix}

\end{document}